\documentclass[11pt]{article}
\usepackage[english]{babel}
\usepackage{csquotes}
\usepackage[letterpaper,top=1.9cm,bottom=1.9cm,left=2.5cm,right=2.5cm,marginparwidth=1.75cm]{geometry}
\usepackage{amsmath}
\usepackage{graphicx}
\usepackage[colorlinks=true, allcolors=blue]{hyperref}
\usepackage{booktabs}
\usepackage{multirow}
\usepackage[table,xcdraw]{xcolor}
\usepackage{setspace}
\usepackage{siunitx}
\usepackage[backend=biber, style=apa]{biblatex}

\DeclareSourcemap{
  \maps[datatype = bibtex]{
    \map{
      \step[notfield = keywords, final]
      \step[fieldsource = doi, final]
      \step[fieldset = url, null]
    }
    \map{
      \step[fieldsource = keywords, notmatch = \regexp{\bprimary\b}, final]
      \step[fieldsource = doi, final]
      \step[fieldset = url, null]
    }
  }
}

\title{
    Comparative analysis of classification schemes\\on major bibliometric platforms:\\A study of Web of Science, Scopus, the Lens, and Dimensions 
}
\author{
    Ophélie Fraisier-Vannier
}
\date{}

\begin{document}

\maketitle

\begin{abstract}
    This study examines the classification schemes of four major bibliometric platforms --~Web of Science, Scopus, Dimensions, and The Lens~-- which collectively index hundreds of millions of scholarly documents.
    As these platforms grow in scale and scope, researchers face increasing complexity when searching and analyzing their vast collections.
    Classification schemes serve as essential tools to navigate these massive datasets, enabling researchers to efficiently filter publications by field or topic.
    However, the substantial differences between platforms' classification approaches --~ranging from expert-curated journal-level categorization to AI-driven document-level classification~-- can significantly impact research outcomes.
    This study provides a comprehensive comparison of each platform's classification methods, structures, and granularity levels.
    By understanding these variations, researchers can make more informed decisions about which platform's classification system best aligns with their specific research objectives, ultimately improving the accuracy and relevance of their bibliometric analyses.
\end{abstract}

\section{Introduction}
    Researchers and scholars frequently encounter the task of constituting bibliometric collections.
    Despite how common this task is, it often raises questions.
    The first questions centers around how to choose which publications will be part of our collection, and which are too far-removed from the topic we wish to focus on to be included.
    Even for long-established scientific fields or topics, these questions have no definitive answer and discussions are often animated concerning what belongs to a given field or not, especially for subjects linked to several disciplines or for front research not yet well established.
    \textcite{Zitt2019} highlight how difficult delineating a scientific field could be, and the various approaches that were available to researchers.
    
    Other questions are more practical: with the multiplication of search engines and databases targeted at researchers and other actors of the research ecosystem, how to choose the best one for the construction of a bibliography?
    Which search fields are the most appropriate to use?
    Here we will focus on the ready-made classifications available on bibliometric platforms.
    Each platform has its strength and weaknesses, and knowing them is essential in order to understand the data retrieved while using them.
    Some publications compared the coverage of some platforms \parencite{gusenbauer_search_2022}, the citation coverage they offer \parencite{martin-martin_google_2020}, or their effectiveness for a specific task, such as performing a systematic reviews in medical informatics \parencite{afraz_literature_2024}.
    However it is very hard to find a consolidated comparison of the various classification schemes they offer, despite them being often used as a shortcut for people looking for a quick overview of a research domain.
    With definitions and scope varying greatly from one scheme to another, it seems important to have a clear overview of what each classification entails and how they are designed.

    In this study, we analyzed 4 bibliometric platforms: Web of Science, Scopus, Dimensions, and The Lens.
    Our aim is to provide an overview of the classification schemes offered by these platforms and how they differ from one another.

\section{Methodology}
    We aimed to demonstrate the diversity of classification schemes across bibliometric platforms.
    We present here the selection process we used to reach the list of classifications presented in this article.

\subsection{Platforms selection}

    Existing bibliometric platforms can be categorized into 2 big families depending on their philosophy regarding the content they index, we will call them `selective platforms' and `non-selective platforms':
    \begin{itemize}
        \item Selective platform:
            This main argument of this type of platform is the quality of the content it contains.
            It indexes only content manually selected by experts.
            The amount of data available on the platform is therefore limited but often manually cleaned and enriched, and the final user can be reassured that the content they are presented with has been validated by experts.
            This philosophy of indexing  was used by the first bibliometric platforms.
        \item Non-selective platform:
            On the other hand, this type of platform’s main force is the quantity of documents ingested.
            It indexes all content available from numerous sources into a unified database.
            The final user sees a much more exhaustive view thanks to the broader coverage, however the level of quality of the records can vary.
            The new generation of bibliometric platforms tend to adopt this approach.
    \end{itemize}
    In addition, some platforms are focused on a specific scientific domain. For example, one of the most well-known bibliometric platforms is PubMed, a platform focused on biomedical scientific publications.

    Given our goal, our final selection criteria for the platforms to consider in the analysis were the following:
    \begin{enumerate}
        \item We wanted both selective and non-selective platforms represented, in order to see if this difference in indexing philosophy had an impact on the classification schemes offered.
        \item We wanted only generalist platforms in order to avoid comparing classification schemes that would not encompass the same scope of scientific research.
    \end{enumerate}

    Based on these criteria, we selected two selective platforms and 2 non-selective platforms: Web of Science (WOS), Scopus, Dimensions, and The Lens.
    An overview of these platforms is provided in \autoref{sec:platforms}.

\subsection{Research classification schemes selection}
    We define here classification as the process of partitioning indexed documents --~representing, metaphorically, the scientific landscape~-- into distinct categories based on specific characteristics, thereby organizing them into coherent subject areas.
    The resulting framework of categories is referred to as the classification scheme.

    Once the platforms were selected, we listed the classification schemes they offered to users to delineate scientific fields or topics.
    In order to stay coherent with the second criterion from our platform selection step, we excluded classifications whose goal was not to map all scientific fields but only specific domains or goals.\footnote{We can cite for example the \emph{Common Scientific Outline (CSO)}, managed by the International Cancer Research Partnership \parencite{icrp_international_2023}, or the \emph{Sustainable Development Goals} defined by the United Nations in 2015, which are available in WOS and Dimensions \parencite{clarivate_incites_2026-1,dimensions_guide_2023}.}
    Additionally, we excluded classifications that were not available on the full platform’s database within a standard research subscription.
    For Dimensions, this concerned a few specific classifications such as the \emph{Broad Research Areas} for example.\footnote{This classification scheme is used by the Australian and New Zealand Standard Research Classification and consists of 4 categories: Basic Science, Clinical Medicine \& Science, Health Services Research, and Public Health.}
    For Web of Science, this meant taking only into account classifications present in a core collection record \parencite{clarivate_core_2025}, and excluding classification schemes available only on specific indexes.
    It also meant excluding additional classifications which can be find in some of Clarivate’s documentation but are in fact offered by InCites --~InCites is another tool commercialized by Clarivate which uses Web of Science data for bibliometric analysis and institutional research evaluation.

    This left us with the following classification schemes used by these platforms (sorted alphabetically):
    \begin{itemize}
        \item `All Science Journal Classification' (available on Scopus and The Lens, where it is called `Subject matter'),
        \item `Citation topics' (available on Web of Science),
        \item `Field of study' (available on The Lens),
        \item `Fields of Research' (available on Dimensions),
        \item `Subject categories' (available on Web of Science),
        \item `Units of Assessment' (available on Dimensions).
    \end{itemize}
    These classifications are presented in more details in \autoref{sec:classifications}.

\section{Presentation of the platforms}\label{sec:platforms}
    The following sections give an overview of the bibliometric platforms we selected for analysis.
    A condensed summary of the platforms and their content is provided in \autoref{tab:pfs}.

    \begin{table}[htbp]
        \caption{Overview of the bibliometric platforms presented~: Web of Science, Scopus, Dimensions, and The Lens -- for-profit organizations are highlighted in yellow and non-profit organizations in pink}
        \label{tab:pfs}
            \begin{center}
        \begin{tabular}{@{}llll@{}}
            \toprule
            Platform & Owner & Launched & Coverage \\
            \midrule
            \begin{minipage}[c]{0.17\linewidth}
                Web of Science\\[3mm]
                \small{\parencite{clarivate_resources_2025}}
            \end{minipage}
             & {\cellcolor{yellow}Clarivate}   & 1997$^1$ &
                \begin{minipage}[c]{0.5\linewidth}\small
                    \begin{itemize}
                        \setlength\itemsep{-0.4em}
                        \item \emph{Core collection}: 95 million records from
                        \begin{itemize}
                            \vspace{-2mm}
                            \setlength\itemsep{-0.4em}
                            \item \num{22619} journals
                            \item \num{157000} books
                            \item \num{314000} conferences
                        \end{itemize}
                        \item Rest of the platform: 235 million records from
                        \begin{itemize}
                            \vspace{-2mm}
                            \setlength\itemsep{-0.4em}
                            \item \num{34865} journals + books + proceedings
                            \item 128 million patents
                            \item 15 million data sets
                        \end{itemize}
                    \end{itemize}
                    \vspace{-1mm}
                \end{minipage}\\
            \midrule
            \begin{minipage}[c]{0.17\linewidth}
                Scopus\\[3mm]
                \small{\parencite{elsevier_scopus_2025}}
            \end{minipage}
             & {\cellcolor{yellow}Elsevier}   & 2004 &
                \begin{minipage}[c]{0.5\linewidth}\small
                    \begin{itemize}
                        \setlength\itemsep{-0.4em}
                        \item 100 million publication records
                        \item \num{28900} journals and conferences
                        \item \num{399000} books
                        \item 2.6 millions preprints
                    \end{itemize}
                    \vspace{-3mm}
                \end{minipage}\\
            \midrule
            \begin{minipage}[c]{0.17\linewidth}
                The Lens\\[3mm]
                \small{\parencite{lens_scholarly_2024}}
            \end{minipage}
             & {\cellcolor{pink}Cambia}   & 2014$^2$ &
                \begin{minipage}[c]{0.5\linewidth}\small
                    \begin{itemize}
                        \setlength\itemsep{-0.4em}
                        \item 284 million scholarly works in total
                        \item 140 million journal articles
                        \item 8 million proceedings
                        \item 7 million books
                        \item 4 million preprints
                        \item 2.1 billion citations
                        \item 7 million datasets
                        \item 160 million patents
                    \end{itemize}
                    \vspace{-3mm}
                \end{minipage}\\
            \midrule
            \begin{minipage}[c]{0.17\linewidth}
                Dimensions\\[3mm]
                \small{\parencite{dimensions_guide_2023}}
            \end{minipage}
             & {\cellcolor{yellow}Digital Science}  & 2018 &
                \begin{minipage}[c]{0.5\linewidth}\small
                    \begin{itemize}
                        \setlength\itemsep{-0.4em}
                        \item 137 million records
                        \item \num{209000} journals, conferences, books, preprints
                        \item 1.7 billion citations
                        \item 7 million grants
                        \item 155 million patents
                        \item 12 million datasets
                        \item \num{800000} clinical trials
                        \item \num{900000} policy documents
                    \end{itemize}
                    \vspace{-3mm}
                \end{minipage}\\
            \bottomrule
        \end{tabular}
    \end{center}
    {\footnotesize $^1$ Some indexes were created as early as 1964 but they were gathered online under the \emph{Web of Science} brand in 1997~\parencite{martin-martin_google_2020}.}
    
    {\footnotesize $^2$ First non-beta release of Lens including scholarly works (0.9), previously available since 2000 as \emph{Patent Lens}.}
    \end{table}

\subsection{Web of Science (WOS)}\label{sec:wos}

    \emph{Web of Science}, often abbreviated as \emph{WOS}, is a product of the for-profit organization \emph{Clarivate}, which commercializes several tools around data analysis in, among others, the academia and government markets~\parencite{clarivate_about_nodate}.\footnote{Clarivate also commercializes the Journal Citation Reports (JCR) which computes journals’ impact factor every year.}
    It is one of the oldest --~it exists in its current form since 1997 but its premises date back from 1964~-- and most used bibliometric platform in research: from 2020 to 2024, it appears in the title or abstract of approximately \num{120000} articles.\footnote{Search done on the platforms Dimensions and The Lens on the 2025-01-23 with the following filters: publication type is `article', publication date is between 2020 and 2024, and `web of science' appears in the title or the abstract.}
    It is mainly known for its \emph{core collection} but it also contains some regional collections, like the Chinese Science Citation Database or the KCI Korean Journal Database, and some more specialized indexes on zoology, botanic, or grants for example.

    \emph{Web of Science core collection} contains 6 scientific productions’ collections, called indexes:
    \begin{itemize}
        \item 4 indexes are dedicated to journals: Science Citation Index Expanded, Social Sciences Citation Index (SSCI), Arts \& Humanities Citation Index (A\&HCI), and Emerging Sources Citation Index (ESCI).
        \item One index is dedicated to conferences: the Conference Proceedings Citation Index, subdivided into 2 parts, Science (CPCI-S) and Social Sciences \& Humanities (CPCI-SSH).
        \item One is dedicated to books: the Book Citation Index, also subdivided into Science (BKCI-S) and Social Sciences \& Humanities (BKCI-SSH).
    \end{itemize}
    
    For each index, a specific set of journals (or respectively conferences or books) is selected by experts and indexed in \emph{WOS}.
    The latest numbers we found were from July 2025, and indicated that \emph{WOS core collection} contained more that 95 million scientific productions with metadata coming from more than \num{22619} journals, \num{157000} books, and \num{314000} conferences (see \autoref{tab:wos} for more details).
    In the rest of the platform, Clarivate declares indexing more than 235 million records from approximately \num{35000} journals, books, or conferences.
    In addition, 128 million patents and more than 15 million data sets can be searched.
    
\begin{table}[htbp]
    \caption{Web of Science core collection overview}
    \label{tab:wos}
    
    \begin{center}
    \begin{tabular}{@{}lrrl@{}}
%    \begin{tabular}{@{}p{7.5cm}rrrp{1.2cm}@{}}
        \toprule
%        Index & Sources$^1$ & Records$^2$ & Subject categories$^3$ & Oldest records 
        Index & Sources$^1$ & Records$^2$ & Oldest records \\
        \midrule
        Science Citation Index Expanded (SCI-EXPANDED)
                            & \num{9478} & \num{62000000}  & 1900\\
        Social Sciences Citation Index (SSCI)
                            & \num{3551} & \num{11000000}  & 1900\\
        Arts \& Humanities Citation Index (A\&HCI)
                            & \num{1824} & \num{6000000}  & 1975\\
        Emerging Sources Citation Index (ESCI)
                            & \num{8507} & \num{5000000}  & 2005\\
        Conference Proceedings Citation Index (CPCI-S, CPCI-SSH)
                            & \num{308000} & \num{13000000}  & 1990\\
        Book Citation Index (BKCI-S, BKCI-SSH)
                            &  & \num{151000}  & 2005\\
%        Science Citation Index Expanded
%                            & \num{9478} & \num{62000000} & 182 & 1900\\
%        Social Sciences Citation Index (SSCI)
%                            & \num{3551} & \num{11000000} & 47 & 1900\\
%        Arts \& Humanities Citation Index (A\&HCI)
%                            & \num{1824} & \num{6000000} & 25 & 1975\\
%        Emerging Sources Citation Index (ESCI)
%                            & \num{8507} & \num{5000000} & 252 & 2005\\
%        Conference Proceedings Citation Index (CPCI-S, CPCI-SSH)
%                            & \num{308000} & \num{13000000} & 254 & 1990\\
%        Book Citation Index (BKCI-S, BKCI-SSH)
%                            &  & \num{151000} & 254 & 2005\\
        \bottomrule
    \end{tabular}
    \end{center}
    \vspace{-3mm}
    \footnotesize{$^1$ Journals, conferences, or books.}\\
    \footnotesize{$^2$ Only records with metadata.}\\
%    \footnotesize{$^3$ WOS core collection has 254 subject categories in total.}
\end{table}

\subsection{Scopus}
    Scopus is a multidisciplinary scholarly database commercialized by the company Elsevier \parencite{elsevier_what_2024}.
    It was launched in 2004 and indexes expert-selected journals, conferences, and books.
    Since Elsevier is also one of the biggest scientific publisher, the selection process is handled by an independent board --~the Scopus Content Selection and Advisory Board (CSAB) comprised of 17 experts~-- who looks at the history, practices, and standing of the proposed journal (resp. conference or book) in their field to decide if it should be indexed in Scopus \parencite{elsevier_scopus_nodate-1}.
    Thailand, Korea, Russian and China also have local boards for matters related to local journals.
    
    Elsevier indicates on their website that in February~2025, more than 100 million records were indexed and available for research, coming from more than \num{28900} journals and conferences\footnote{The full list of Scopus source titles can be downloaded here: \url{https://www.elsevier.com/products/scopus/content\#4-titles-on-scopus}}, \num{399000} books, and 2.6 millions preprints \parencite{elsevier_scopus_2025,elsevier_scopus_nodate-1}.

\subsection{Dimensions}
    Dimensions was launched in 2018 by the technology company Digital Science.
    The aim of the platform is to aggregate and provide links between different types of scholarly data: publications, grants, datasets, patents, clinical guidelines. In this study, we will only focus on scientific publications.
    Contrary to WOS and Scopus, Dimensions does not have a selective approach to scientific outlets.
    They declare indexing ``all source titles covered by our publication data sources such as Crossref, PubMed, Europe PubMed Central, arXiv and partnerships with 130 publishers".
    This approach allowed them to index close to 140 million scientific records along with close to 2 billion citations, in addition to several million patents, datasets, and other documents related to research \parencite{dimensions_guide_2023, dimensions_why}

    Some analyses done when the platform was launched showed that it could be an alternative to the more established and selective platforms~-- \textcite{thelwall_dimensions_2018} found for example that 97\% of the articles indexed in Scopus were also present in Dimensions.

\subsection{The Lens}
    The Lens is a platform owned by the non-profit organization Cambia.
    It was initially launched in 2000 under the name \emph{Patent Lens} and focused on patents.
    In 2013, the current version of \emph{The Lens} was launched, it aggregates both scholarly publications and patents and links them when possible.
    
    Similarly to \emph{Dimensions}, \emph{the Lens} does not have a selective approach to scientific publications.
    According to the information visible on their platform\footnote{All numbers coming from the platform (\url{https://www.lens.org/lens/search/scholar/structured}) for this section were collected on the 2025-01-30.}, they collect publications from the following sources: Microsoft Academic (203 million publications), Crossref (165 million publications), OpenAlex (67 million publications), and PubMed (38 million publications), with a large overlap between these four sources.
    In total, their website indicates that the platform contains more than 284 million scholarly works connected by more than 2 billion citations, including 140 million journal articles and 8 million conference proceedings.\footnote{We did not find any public numbers available about the number of journals present on the platform.}

\section{Presentation of the classification schemes}\label{sec:classifications}

    This section presents the different classification schemes selected.
    It provides an overview of each scheme, as well as more hands-on information such as the number of categories it contains, on which platform it is available, and how are documents assigned to a category.
    \autoref{tab:classif_per_platform} gives a summarized view of each classification scheme.
    
    \begin{table}[htbp]
        \caption{
            Availability of classification per bibliometric platform. 
            \colorbox[HTML]{CBCEFB}{J} indicates the classification is done by experts at journal-level, \colorbox[HTML]{FFFFC7}{D} indicates the classification is done by artificial intelligence at document-level.
            $^+$ means that the same document can be assigned to several categories of the classification simultaneously.
        }
        \label{tab:classif_per_platform}
        
    \begin{center}
    \begin{tabular}{@{}lr@{}>{ }lcccc@{}}
\toprule
    & \multicolumn{2}{c}{Number of} & \multicolumn{4}{c}{Bibliometric platform} \\
    \cmidrule{4-7}
    Classification & \multicolumn{2}{c}{categories} & WOS & Scopus & Dimensions & Lens \\
\midrule
    \multirow{3}{*}{All Science Journal Classification}
        & 4
        & main areas
        &
        & \cellcolor[HTML]{CBCEFB}
        &
        & \cellcolor[HTML]{CBCEFB}
        \\
%    All Science Journal Classification
        & 26
        & subject areas
        &
        & \cellcolor[HTML]{CBCEFB}
        &
        & \cellcolor[HTML]{CBCEFB}
        \\
%    All Science Journal Classification
        & 361
        & themes
        &
        & \cellcolor[HTML]{CBCEFB}\multirow{-3}{*}{J$^+$}
        &
        & \cellcolor[HTML]{CBCEFB}\multirow{-3}{*}{J$^{+,1}$}
        \\
    &&&&&&\\[-2mm]
    \multirow{3}{*}{Citation topics}
        & 10
        & macro-topics% (TMAC)
        & \cellcolor[HTML]{FFFFC7}
        &
        &
        &
        \\
%    Citation topics
        & 326
        & meso-topics% (TMSO)
        & \cellcolor[HTML]{FFFFC7}
        &
        &
        &
        \\
%    Citation topics
        & 2493
        & micro-topics% (TMIC)
        & \cellcolor[HTML]{FFFFC7}\multirow{-3}{*}{D}
        &
        &
        &
        \\
    &&&&&&\\[-2mm]
    \multirow{4}{*}{Field of Study$^2$}
        & 4
        & domains
        &
        &
        &
        &
        \cellcolor[HTML]{FFFFC7}
        \\
%    Field of Study
        & 26
        & fields
        &
        &
        &
        &
        \cellcolor[HTML]{FFFFC7}
        \\
%    Field of Study
        & 252
        & subfields
        &
        &
        &
        &
        \cellcolor[HTML]{FFFFC7}
        \\
%    Field of Study
        & 4516
        & topics
        &
        &
        &
        &
        \cellcolor[HTML]{FFFFC7}\multirow{-4}{*}{D$^+$}
        \\
    &&&&&&\\[-2mm]
    \multirow{3}{*}{Fields of Research}
        & 23
        & divisions
        &
        &
        &
        \cellcolor[HTML]{FFFFC7}
        & 
        \\
%    Fields of Research 
        & 213
        & groups
        &
        &
        &
        \cellcolor[HTML]{FFFFC7}
        & 
        \\
%    Fields of Research 
        & 1967
        & fields
        &
        &
        &
        \cellcolor[HTML]{FFFFC7}\multirow{-3}{*}{D$^{+,3}$}
        & 
        \\
    &&&&&&\\[-2mm]
    Subject categories
        & 254
        &
        & \cellcolor[HTML]{CBCEFB}J$^+$ 
        &
        &
        &
        \\
    &&&&&&\\[-2mm]
%    Sustainable Development Goals
%        & 17
%        &
%        & \cellcolor[HTML]{FFFFC7}D$^+$ 
%        &
%        & \cellcolor[HTML]{FFFFC7}D$^+$
%        &
%        \\
    &&&&&&\\[-2mm]
    Units of Assessment
        & 34
        &
        &
        &
        & \cellcolor[HTML]{FFFFC7}D$^+$
        &
        \\
\bottomrule
    \end{tabular}
    \end{center}
    \vspace{-3mm}
    \footnotesize{$^1$ Classification called \emph{Subject matter} in Lens.}\\
    \footnotesize{$^2$ The \emph{field of study} classification presented in Lens is a concatenation of the now-retired Microsoft Academic topics, used before 2021, and OpenAlex topics, used for newly published articles. For clarity-sake, we present here only the number of categories present in OpenAlex topics since it is the classification currently used by the platform for new publications.}\\
    \footnotesize{$^3$ Only 22 divisions and 171 groups are available in Dimensions: the "Indigenous Studies" division was filtered out, as well as all the "Other" groups (generic groups gathering the publications that cannot be classified in any other group of the division, such as \emph{Other biological sciences} for example).
    The field granularity is also not available.
    An additional step of machine-learning classification can be done at journal-level if not enough text is available at article-level, see \autoref{sec:dimensions_FoR} for more information.}
    \end{table}

\subsection{All Science Journal Classification (ASJC)}\label{sec:asjc}

    The `All Science Journal Classification' (ASJC) was created by Scopus to categorize the journals indexed on the platform.
    It consists of 361 themes, each assigned to a name and a 4-digits code, divided into 26 subject area classifications across 4 main areas (\emph{Physical Sciences}, \emph{Health Sciences}, \emph{Social Sciences}, \emph{Life Sciences}) plus a \emph{multidisciplinary} category (see \autoref{tab:asjc}).
    When a journal or conference is first selected to be indexed in Scopus, experts assign it to one or several subjects areas based on its content \parencite{elsevier_what_2024}.

    \begin{table}[btp]
        \caption{Scopus All Science Journal Classification (ASJC) categories organization}\label{tab:asjc}
                \begin{center}
        \begin{tabular}{@{}lll@{}}
            \toprule
            Main Area        &	Subject Area   & ASJC themes  \\
            \midrule
            Physical Sciences   &	Chemical Engineering                            & 1500 to 1508 \\
                                &   Chemistry                                       & 1600 to 1607 \\
                                &   Computer Science                                & 1700 to 1712 \\
                                &   Earth and Planetary Sciences                    & 1900 to 1913 \\
                                &   Energy                                          & 2100 to 2105 \\
                                &   Engineering                                     & 2200 to 2216 \\
                                &   Environmental Science                           & 2300 to 2312 \\
                                &   Material Science                                & 2500 to 2508 \\
                                &   Mathematics                                     & 2600 to 2614 \\
                                &   Physics and Astronomy                           & 3100 to 3110 \\
            \midrule
            Health Sciences	    &   Medicine                                        & 2700 to 2748 \\
                                &   Nursing                                         & 2900 to 2923 \\
                                &   Veterinary                                      & 3400 to 3404 \\
                                &   Dentistry                                       & 3500 to 3506 \\
                                &   Health Professions                              & 3600 to 3616 \\
            \midrule
            Social Sciences	    &   Arts and Humanities                             & 1200 to 1213 \\
                                &    Business, Management and Accounting            & 1400 to 1410 \\
                                &    Decision Sciences                              & 1800 to 1804 \\
                                &    Economics, Econometrics and Finance            & 2000 to 2003 \\
                                &    Psychology                                     & 3200 to 3207 \\
                                &    Social Sciences                                & 3300 to 3322 \\
            \midrule
            Life Sciences	    &    Agricultural and Biological Sciences           & 1100 to 1111 \\
                                &    Biochemistry, Genetics and Molecular Biology   & 1300 to 1315 \\
                                &    Immunology and Microbiology                    & 2400 to 2406 \\
                                &    Neuroscience                                   & 2800 to 2809 \\
                                &    Pharmacology, Toxicology and Pharmaceutics     & 3000 to 3005 \\
            \midrule
            \multicolumn{2}{@{}l}{Multidisciplinary}                                   & 1000         \\
%            \midrule
%            \emph{Total}    & \hfill\emph{27}    & \hfill\emph{362}   \\
            \bottomrule
        \end{tabular}
        \end{center}
    \end{table}

    This classification is also available on The Lens \parencite{lens_scholar_2024}.
    The platform used to retrieve the ASJC categories assigned to sources using Crossref but the field containing this information was deprecated in 2024.
    Despite this, The Lens still offers this classification thanks to a mapping they maintain internally.

\subsection{Citation topics}\label{sec:wos_citation_topics}
    In scholarly communication, a citation is the act of formally acknowledging an external source through a bibliographic reference.
    It also represents the link between cited works and their original source.

    `Citation topics' are computed by the \emph{WOS} platform.
    They are the result of a clustering algorithm based on citation links \parencite{clarivate_incites_2026}.
    This clustering is done at document-level and organized into 3 hierarchical levels: 10 macro-topics (also called TMAC in \emph{WOS}),  326 meso-topics (TMSO), and more than \num{2000} micro-topics (TMIC, see \autoref{tab:wos_citation_topics} for more details).
    Each topic has a name\footnote{Macro- and meso-topics are labeled by hand, while micro-topics are labeled by automatically inferring a representative keyword based on the concerned documents.} and a permanent numerical prefix: \texttt{A.B.C} identifies a micro-topic, \texttt{A.B} the corresponding meso-topic, and \texttt{A} the macro-topic above.
    Each document can only be assigned to a single topic, and some documents do not have any citation topic.\footnote{\emph{WOS} estimates that it concerns approximately 10\% of articles and reviews published since 1980 indexed on their platform.}

    An update of the clusters is done every year.
    This update retains the macro and meso structure but can add new micro-topics and potentially move some documents between micro-topics.

    \begin{table}[btp]
        \caption{Summary of Web of Science citation topics}
        \label{tab:wos_citation_topics}
        {
            \centering
            \small
            
\begin{tabular}{@{}rp{0.23\linewidth}>{\footnotesize}p{0.35\linewidth}rr@{}}
\toprule
\multicolumn{2}{l}{{Macro-topic}}                                                     & Examples of meso-topics & \# Meso-topics & \# Micro-topics \\
\midrule
{1}
    & {Clinical \& Life Sciences}
    & 1.5 Neuroscience, 1.6 Immunology, 1.7 Neuroscanning
    & {132} 
    & {946}                  
    \\[7mm]
{2}  
    & {Chemistry}      
    & 2.1 Synthesis, 2.15 Physical Chemistry, 2.22 Inorganic \& Nuclear Chemistry
    & {37}         
    & {265}           
    \\[7mm]
{3} 
    & {Agriculture, Environment \& Ecology}     
    & 3.87 Biobased Materials \& Bioenergy, 3.91 Contamination \& Phytoremediation, 3.97 Plant Pathology
    & {24}                    
    & {247}            
    \\[10mm]
{4}  
    & {Electrical Engineering, Electronics \& Computer Science} 
    & 4.47 Software Engineering, 4.48 Information Retrieval \& Knowledge Systems, 4.58 Wireless Technology
    & {26}          
    & {231}         
    \\[10mm]
{5} 
    & {Physics}       
    & 5.107 Laser Science, 5.131 Space \& Plasma Physics, 5.135 Nuclear Physics
    & {27}            
    & {167}        
    \\[7mm]
{6} 
    & {Social Sciences}    
    & 6.73 Social Psychology, 6.86 Human Geography, 6.110 Law
    & {28}          
    & {240}     
    \\[7mm]
{7} 
    & {Engineering \& Materials Science}   
    & 7.260 Nuclear Engineering, 7.262 Explosives, 7.272 Electrical - Solder \& Connections
    & {20}       
    & {139}     
    \\[7mm]
{8}  
    & {Earth Sciences}      
    & 8.93 Archaeology, 8.124 Environmental Sciences, 8.305 Paleontology
    & {12}           
    & {69}          
    \\[7mm]
{9}  
    & {Mathematics}    
    & 9.28 Pure Maths, 9.162 Numerical Methods, 9.280 Algebra \& Topology
    & {8}        
    & {67}       
    \\[7mm]
{10} 
    & {Arts \& Humanities}  
    & 10.99 Literary Theory, 10.126 Philosophy, 10.144 Modern History
    & {12}              
    & {122}         
    \\
\midrule
\multicolumn{3}{r}{{\itshape Total}}
    & \textit{326}
    & \textit{2493}   
    \\
\bottomrule
\end{tabular}
        }
    \end{table}

\subsection{Field of Study}
    The Lens offers for each publication a classification done via a machine-learning model \parencite{lens_scholar_2024}.
    The classification model was initially based on the \emph{Fields of Study} from Microsoft Academic.
    Since the retirement of Microsoft Academic in 2021, The Lens uses the Concepts taxonomy from OpenAlex.

    OpenAlex topics are based on the publications’ citation network and organized into 4 hierarchical levels \parencite{openalex_openalex_2024}: 4 domains (1-digit codes), 26 fields (2-digits codes), 252 subfields (4-digits codes), and 4516 topics (5-digits codes).
    Domains, fields, and subfields are based on Scopus ASJC categories (see \autoref{sec:asjc} for more information of this classification scheme).
    Each topic is also assigned a list of keywords describing its content.
    A single article can be assigned to several topics.
    
    \begin{table}[htbp]
        \caption{Hierarchical levels of OpenAlex topics}\label{tab:openalex_hierarchy}
        \centering
        \begin{tabular}{@{}lrll@{}}
            \toprule
            Hierarchical level  & Number of categories  & \multicolumn{2}{l}{\small{Example}}  \\
            \midrule
            Domain      &   4       & \small 1     & \small Life Sciences  \\
            Field       &   26      & \small 11    & \small Agricultural and Biological Sciences \\
            Subfield    &   252     & \small 1106  & \small Food Science \\
            Topic       &   4516    & \small 10650 & \small Food Drying and Modeling \\
            \bottomrule
        \end{tabular}
    \end{table}

\subsection{Fields of Research (ANZSRC 2020)}\label{sec:dimensions_FoR}
    This classification is based on the \emph{Australian and New Zealand Standard Research Classification}  or ANZSRC \parencite{australian_bureau_of_statistics_australian_2020}.
    The latest version of the classification was published in 2020 and provides 3 classifications for research and R\&D: Type of Activity (ToA), Fields of Research (FoR), Socio-Economic Objectives (SEO).
    The \emph{Fields of Research} are organized in 3 hierarchical levels: 23 divisions (2-digits codes), 213 groups (4-digits codes), and 1967 fields for the finest level (6-digits codes) (see \autoref{tab:anzsrc_hierarchy} and \autoref{tab:anzsrc} for more details and examples).

    \begin{table}[htbp]
        \caption{Hierarchical levels of the Australian and New Zealand Standard Research Classification (ANZSRC 2020) Fields of Research classification scheme}\label{tab:anzsrc_hierarchy}
        \centering
        \begin{tabular}{@{}lrll@{}}
            \toprule
            {Hierarchical level}  
            & {Number of categories}
            & \multicolumn{2}{l}{\small{Example}}    \\
            \midrule
            Division    &   23 & \small 31      & \small Biological sciences \\
            Group       &  213 & \small 3103    & \small Ecology  \\
            Field       & 1967 & \small 310301  & \small Behavioural ecology  \\ 
            \bottomrule
        \end{tabular}
    \end{table}

    \begin{table}[p]
        \caption{Categories of the Australian and New Zealand Standard Research Classification (ANZSRC 2020) Fields of Research}\label{tab:anzsrc}
        {\small
        
\begin{tabular}{@{}p{0.3\linewidth}>{\footnotesize}p{0.5\linewidth}rr@{}}
\toprule
Division & Examples of groups & \# groups & \# fields \\
\midrule
Agricultural, veterinary and food sciences~(30)
    & Agricultural biotechnology~(3001), Agriculture, land and farm management~(3002)
    & 10
    & 95               \\[5mm]
Biological sciences~(31)
    & Biochemistry and cell biology~(3101), Bioinformatics and computational biology~(3102)
    & 10
    & 98               \\[5mm]
Biomedical and clinical sciences~(32)
    & Cardiovascular medicine and haematology~(3201), Clinical sciences~(3202)
    & 16
    & 129              \\[5mm]
Built environment and design~(33)
    & Architecture~(3301), Building~(3302)
    & 5
    & 51               \\[5mm]
Chemical sciences~(34)
    & Analytical chemistry~(3401), Inorganic chemistry~(3402)
    & 8
    & 63               \\[5mm]
Commerce, management, tourism and services~(35)
    & Accounting, auditing and accountability~(3501), Banking, finance and investment~(3502)
    & 10
    & 91               \\[5mm]
Creative arts and writing~(36)
    & Art history, theory and criticism~(3601), Creative and professional writing~(3602)
    & 7
    & 35               \\[5mm]
Earth sciences~(37)
    & Atmospheric sciences~(3701), Climate change science~(3702)
    & 10
    & 65               \\[5mm]
Economics~(38)
    & Applied economics~(3801), Econometrics~(3802)
    & 4
    & 35               \\[5mm]
Education~(39)
    & Curriculum and pedagogy~(3901), Education policy, sociology and philosophy~(3902)
    & 5
    & 43               \\[5mm]
Engineering~(40)
    & Aerospace engineering~(4001), Automotive engineering~(4002)
    & 20
    & 196              \\[5mm]
Environmental sciences~(41)
    & Climate change impacts and adaptation~(4101), Ecological applications~(4102)
    & 7
    & 38               \\[5mm]
Health sciences~(42)
    & Allied health and rehabilitation science~(4201), Epidemiology~(4202)
    & 9
    & 71               \\[5mm]
History, heritage and archaeology~(43)
    & Archaeology~(4301), Heritage, archive and museum studies~(4302)
    & 4
    & 44               \\[5mm]
Human society~(44)
    & Anthropology~(4401), Criminology~(4402)
    & 11
    & 116              \\[5mm]
Indigenous studies~(45)
    & Aboriginal and Torres Strait Islander culture, language and history~(4501), Aboriginal and Torres Strait Islander education~(4502)
    & 20
    & 320              \\[5mm]
Information and computing sciences~(46)
    & Applied computing~(4601), Artificial intelligence~(4602)
    & 14
    & 132              \\[5mm]
Language, communication and culture~(47)
    & Communication and media studies~(4701), Cultural studies~(4702)
    & 6
    & 91               \\[5mm]
Law and legal studies~(48)
    & Commercial law~(4801), Environmental and resources law~(4802)
    & 8
    & 62               \\[5mm]
Mathematical sciences~(49)
    & Applied mathematics~(4901), Mathematical physics~(4902)
    & 6
    & 48               \\[5mm]
Philosophy and religious studies~(50)
    & Applied ethics~(5001), History and philosophy of specific fields~(5002)
    & 6
    & 50               \\[5mm]
Physical sciences~(51)
    & Astronomical sciences~(5101), Atomic, molecular and optical physics~(5102)
    & 11
    & 58               \\[5mm]
Psychology~(52)
    & Applied and developmental psychology~(5201), Biological psychology~(5202)
    & 6
    & 36       \\
\bottomrule
\end{tabular}
        }
    \end{table}

    The FoR classification is available in Dimensions thanks to a machine-learning approach applied at document-level using the title and abstract \parencite{dimensions_guide_2023}.
    When articles contain too little text to classify them effectively, a journal-level classification is used instead \parencite{dimensions_what_2025}.
    The categories available differ slightly from the standard presented above.
    Only division and group levels are available for filtering publications, meaning that users cannot use the finest granularity of the classification (the fields).
    Some categories were also not implemented: the `Indigenous Studies' division was filtered out, as well as all the `Other' groups.\footnote{Each division has a generic group to gather the publications that cannot be classified in any group, for example the group \emph{Other biological sciences (3299)} for the \emph{Biological sciences} division (31), or the group \emph{Other psychology (5299)} for the \emph{Psychology} division (52).}

\subsection{Subject categories}
    Each record in the \emph{WOS core collection} is assigned to at least one of the 254 available subject categories, such as `Agronomy', `Management' or `Political Sciences' (see \autoref{tab:wos_subject_cat} for the full list).
    The assignation is done \textbf{at source-level} when a journal (resp. a conference or a book) is first indexed in \emph{WOS} Journal Citation Report (JCR): it is assigned by \emph{WOS} experts to one or several subject categories.
    These subject categories are then propagated to every entry linked to this source \parencite{clarivate_web_2025}.

    Some research fields can be considered too vast to be in a single category, in that case they encompass several subject categories.
    It is for example the case of the engineering fields which is directly linked to 14 subject categories\footnote{`Engineering, Aerospace', `Engineering, Biomedical', `Engineering, Chemical', `Engineering, Civil', `Engineering, Electrical \& Electronic', ` Engineering, Environmental', `Engineering, Geological', `Engineering, Industrial', `Engineering, Manufacturing', `Engineering, Marine', `Engineering, Mechanical', `Engineering, Multidisciplinary', `Engineering, Ocean', `Engineering, Petroleum'} or literature, directly linked to 7 subject categories\footnote{`Literature', `Literature, African, Australian, Canadian', `Literature, American', `Literature, British Isles', `Literature, German, Dutch, Scandinavian', `Literature, Slavic'}.
    On the other hand, some fields are contained is a single subject category, such as `Plant Sciences' or `Economics'.

    \begin{table}[p]
        \caption{List of the 254 Web of Science subject categories}
        \label{tab:wos_subject_cat}
        {
            \centering
            \tiny
            
\begin{tabular}{lllll}
\cmidrule{1-1}
\cmidrule{3-3}
\cmidrule{5-5}\\
Subject categories (part 1/3)                    &  & Subject categories (part 2/3)                  &  & Subject categories (part 3/3)                  \\
\cmidrule{1-1}
\cmidrule{3-3}
\cmidrule{5-5}\\
Acoustics                                        &  & Ergonomics                                     &  & Nuclear Science \& Technology                  \\
Agricultural Economics \& Policy                 &  & Ethics                                         &  & Nursing                                        \\
Agricultural Engineering                         &  & Ethnic Studies                                 &  & Nutrition \& Dietetics                         \\
Agriculture, Dairy \& Animal Science             &  & Evolutionary Biology                           &  & Obstetrics \& Gynecology                       \\
Agriculture, Multidisciplinary                   &  & Family Studies                                 &  & Oceanography                                   \\
Agronomy                                         &  & Film, Radio, Television                        &  & Oncology                                       \\
Allergy                                          &  & Fisheries                                      &  & Operations Research \& Management Science      \\
Anatomy \& Morphology                            &  & Folklore                                       &  & Ophthalmology                                  \\
Andrology                                        &  & Food Science \& Technology                     &  & Optics                                         \\
Anesthesiology                                   &  & Forestry                                       &  & Ornithology                                    \\
Anthropology                                     &  & Gastroenterology \& Hepatology                 &  & Orthopedics                                    \\
Archaeology                                      &  & Genetics \& Heredity                           &  & Otorhinolaryngology                            \\
Architecture                                     &  & Geochemistry \& Geophysics                     &  & Paleontology                                   \\
Area Studies                                     &  & Geography                                      &  & Parasitology                                   \\
Art                                              &  & Geography, Physical                            &  & Pathology                                      \\
Asian Studies                                    &  & Geology                                        &  & Pediatrics                                     \\
Astronomy \& Astrophysics                        &  & Geosciences, Multidisciplinary                 &  & Peripheral Vascular Disease                    \\
Audiology \& Speech-Language Pathology           &  & Geriatrics \& Gerontology                      &  & Pharmacology \& Pharmacy                       \\
Automation \& Control Systems                    &  & Gerontology                                    &  & Philosophy                                     \\
Behavioral Sciences                              &  & Green \& Sustainable Science \& Technology     &  & Physics, Applied                               \\
Biochemical Research Methods                     &  & Health Care Sciences \& Services               &  & Physics, Atomic, Molecular \& Chemical         \\
Biochemistry \& Molecular Biology                &  & Health Policy \& Services                      &  & Physics, Condensed Matter                      \\
Biodiversity Conservation                        &  & Hematology                                     &  & Physics, Fluids \& Plasmas                     \\
Biology                                          &  & History                                        &  & Physics, Mathematical                          \\
Biophysics                                       &  & History \& Philosophy of Science               &  & Physics, Multidisciplinary                     \\
Biotechnology \& Applied Microbiology            &  & History of Social Sciences                     &  & Physics, Nuclear                               \\
Business                                         &  & Horticulture                                   &  & Physics, Particles \& Fields                   \\
Business, Finance                                &  & Hospitality, Leisure, Sport \& Tourism         &  & Physiology                                     \\
Cardiac \& Cardiovascular Systems                &  & Humanities, Multidisciplinary                  &  & Plant Sciences                                 \\
Cell Biology                                     &  & Imaging Science \& Photographic Technology     &  & Poetry                                         \\
Cell \& Tissue Engineering                       &  & Immunology                                     &  & Political Science                              \\
Chemistry, Analytical                            &  & Industrial Relations \& Labor                  &  & Polymer Science                                \\
Chemistry, Applied                               &  & Infectious Diseases                            &  & Primary Health Care                            \\
Chemistry, Inorganic \& Nuclear                  &  & Information Science \& Library Science         &  & Psychiatry                                     \\
Chemistry, Medicinal                             &  & Instruments \& Instrumentation                 &  & Psychology                                     \\
Chemistry, Multidisciplinary                     &  & Integrative \& Complementary Medicine          &  & Psychology, Applied                            \\
Chemistry, Organic                               &  & International Relations                        &  & Psychology, Biological                         \\
Chemistry, Physical                              &  & Language \& Linguistics                        &  & Psychology, Clinical                           \\
Classics                                         &  & Law                                            &  & Psychology, Developmental                      \\
Clinical Neurology                               &  & Limnology                                      &  & Psychology, Educational                        \\
Communication                                    &  & Linguistics                                    &  & Psychology, Experimental                       \\
Computer Science, Artificial Intelligence        &  & Literary Reviews                               &  & Psychology, Mathematical                       \\
Computer Science, Cybernetics                    &  & Literary Theory \& Criticism                   &  & Psychology, Multidisciplinary                  \\
Computer Science, Hardware \& Architecture       &  & Literature                                     &  & Psychology, Psychoanalysis                     \\
Computer Science, Information Systems            &  & Literature, African, Australian, Canadian      &  & Psychology, Social                             \\
Computer Science, Interdisciplinary Applications &  & Literature, American                           &  & Public Administration                          \\
Computer Science, Software Engineering           &  & Literature, British Isles                      &  & Public, Environmental \& Occupational Health   \\
Computer Science, Theory \& Methods              &  & Literature, German, Dutch, Scandinavian        &  & Quantum Science \& Technology                  \\
Construction \& Building Technology              &  & Literature, Romance                            &  & Radiology, Nuclear Medicine \& Medical Imaging \\
Criminology \& Penology                          &  & Literature, Slavic                             &  & Regional \& Urban Planning                     \\
Critical Care Medicine                           &  & Logic                                          &  & Rehabilitation                                 \\
Crystallography                                  &  & Management                                     &  & Religion                                       \\
Cultural Studies                                 &  & Marine \& Freshwater Biology                   &  & Remote Sensing                                 \\
Dance                                            &  & Materials Science, Biomaterials                &  & Reproductive Biology                           \\
Demography                                       &  & Materials Science, Ceramics                    &  & Respiratory System                             \\
Dentistry, Oral Surgery \& Medicine              &  & Materials Science, Characterization \& Testing &  & Rheumatology                                   \\
Dermatology                                      &  & Materials Science, Coatings \& Films           &  & Robotics                                       \\
Development Studies                              &  & Materials Science, Composites                  &  & Social Issues                                  \\
Developmental Biology                            &  & Materials Science, Multidisciplinary           &  & Social Sciences, Biomedical                    \\
Ecology                                          &  & Materials Science, Paper \& Wood               &  & Social Sciences, Interdisciplinary             \\
Economics                                        &  & Materials Science, Textiles                    &  & Social Sciences, Mathematical Methods          \\
Education \& Educational Research                &  & Mathematical \& Computational Biology          &  & Social Work                                    \\
Education, Scientific Disciplines                &  & Mathematics                                    &  & Sociology                                      \\
Education, Special                               &  & Mathematics, Applied                           &  & Soil Science                                   \\
Electrochemistry                                 &  & Mathematics, Interdisciplinary Applications    &  & Spectroscopy                                   \\
Emergency Medicine                               &  & Mechanics                                      &  & Sport Sciences                                 \\
Endocrinology \& Metabolism                      &  & Medical Ethics                                 &  & Statistics \& Probability                      \\
Energy \& Fuels                                  &  & Medical Informatics                            &  & Substance Abuse                                \\
Engineering, Aerospace                           &  & Medical Laboratory Technology                  &  & Surgery                                        \\
Engineering, Biomedical                          &  & Medicine, General \& Internal                  &  & Telecommunications                             \\
Engineering, Chemical                            &  & Medicine, Legal                                &  & Theater                                        \\
Engineering, Civil                               &  & Medicine, Research \& Experimental             &  & Thermodynamics                                 \\
Engineering, Electrical \& Electronic            &  & Medieval \& Renaissance Studies                &  & Toxicology                                     \\
Engineering, Environmental                       &  & Metallurgy \& Metallurgical Engineering        &  & Transplantation                                \\
Engineering, Geological                          &  & Meteorology \& Atmospheric Sciences            &  & Transportation                                 \\
Engineering, Industrial                          &  & Microbiology                                   &  & Transportation Science \& Technology           \\
Engineering, Manufacturing                       &  & Microscopy                                     &  & Tropical Medicine                              \\
Engineering, Marine                              &  & Mineralogy                                     &  & Urban Studies                                  \\
Engineering, Mechanical                          &  & Mining \& Mineral Processing                   &  & Urology \& Nephrology                          \\
Engineering, Multidisciplinary                   &  & Multidisciplinary Sciences                     &  & Veterinary Sciences                            \\
Engineering, Ocean                               &  & Music                                          &  & Virology                                       \\
Engineering, Petroleum                           &  & Mycology                                       &  & Water Resources                                \\
Entomology                                       &  & Nanoscience \& Nanotechnology                  &  & Women's Studies                                \\
Environmental Sciences                           &  & Neuroimaging                                   &  & Zoology                                        \\
Environmental Studies                            &  & Neurosciences                                  &  &                                               \\
\cmidrule{1-1}
\cmidrule{3-3}
\cmidrule{5-5}\\
\end{tabular}

        }
    \end{table}

\subsection{Units of Assessment (UoA)}
    \emph{Units of Assessment} are part of the UK Research Excellence Framework \parencite{ref_2021_uk_2021}. They are used by higher education institutions making submissions to identify their research fields and are composed of 34 categories:

{\small
    \begin{minipage}[t]{0.49\textwidth}
        \begin{enumerate}
            \setlength\itemsep{0em}
            \item Clinical Medicine                                  
            \item Public Health, Health Services and Primary Care    
            \item Allied Health Professions, Dentistry, Nursing and Pharmacy
            \item Psychology, Psychiatry and Neuroscience
            \item Biological Sciences                                      Agriculture, Food and Veterinary Sciences
            \item Earth Systems and Environmental Sciences
            \item Chemistry
            \item Physics
            \item Mathematical Sciences
            \item Computer Science and Informatics
            \item Engineering
            \item Architecture, Built Environment and Planning  
            \item Geography and Environmental Studies
            \item Archaeology
            \item Economics and Econometrics
            \item Business and Management Studies
            \item Law 
        \end{enumerate}
    \end{minipage}
    \begin{minipage}[t]{0.49\textwidth}
        \begin{enumerate}
            \setcounter{enumi}{18}
            \setlength\itemsep{0em}
            \item Politics and International Studies  
            \item Social Work and Social Policy                      
            \item Sociology 
            \item Anthropology and Development Studies   
            \item Education 
            \item Sport and Exercise Sciences, Leisure and Tourism
            \item Area Studies
            \item Modern Languages and Linguistics 
            \item English Language and Literature
            \item History
            \item Classics 
            \item Philosophy
            \item Theology and Religious Studies 
            \item Art and Design: History, Practice and Theory
            \item Music, Drama, Dance, Performing Arts, Film and Screen Studies
            \item Communication, Cultural and Media Studies, Library and Information Management
        \end{enumerate}
    \end{minipage}
}

    This classification scheme is available in Dimensions, where it is applied at document level via a machine-learning approach using the title and abstract \parencite{dimensions_guide_2023}.

\section{Discussion}

\subsection{Sources and actors of classification}
We can detect two main sources of content guiding the articles' classification:
\begin{itemize}
    \item The first is the textual content of the publications;
    \item The other is the citation network around them, which is used to place them in the scientific landscape.
\end{itemize}

Although it is complicated to have details on how these elements are used in the classification process, we can assume that manual expert-based classification happening at journal-level (see \autoref{tab:classif_per_platform} for more details) rely mainly on the content common to the publications published in the same venue.
Among the classification done via machine-learning at document-level, WOS topics and The Lens’ field of study (via OpenAlex topics) are based on the publications citations, whereas in Dimensions the titles and abstracts are the only elements used.

\subsection{Several granularities available}
We can see 4 levels of granularity in the classification schemes presented here, that we will try to name according to the breadth of their focus.

\paragraph{Main areas of science} This granularity consists of only four very high-level areas: `Physical sciences', `Health sciences', `Social sciences', and `Life sciences'.
It is only present in the ASJC and The Lens’ field of study (the latter being based on the former).

\paragraph{Scientific disciplines} We designate here classification levels having no more than a few dozen categories: subject areas from ASJC/The Lens’ field of study (26 categories) and divisions from the ANZSRC fields of research (FoR) available in Dimensions (23 categories).
This granularity tries to encompass in a single category what we could call `disciplines': for example `Computer science', `Psychology', `Engineering', or `Philosophy and religious studies'.
WOS macro-topics (10 categories) fall somewhere in between this level of granularity and the previous one, having both very high-level categories (`Arts \& Humanities' or `Social Sciences') and more disciplines-related ones (`Chemistry' or `Mathematics').

\paragraph{Scientific subfields} We count here classification levels having several hundreds categories, allowing for the selection of only part of a scientific discipline: the themes in ASJC (361 categories), WOS meso-topics (326 categories), The Lens’ subfields of study (252 categories), and FoR groups (213 categories).
We find here values like `Software engineering', `Food science', or `Cultural studies'.

\paragraph{Narrow topics} Finally, we have the finest classification levels, with several thousand values available to filter articles: WOS micro-topics (approximately 2500 categories), The Lens’ field of study topics (more than 4500 categories), and the FoR fields (almost 2000 categories)
We can here select very precise topics, such as `Food drying and modeling' or `Behavioural ecology'.

\paragraph{}
It is important to note that this generalization is far from perfect.
Since the various classification schemes are not aligned, a same label can sit at two very different granularities depending on the chosen classification.
For example, `Health sciences' is one of ASJC four main areas, but is a division in the Fields or Research (so 1 category out of 26).

This introduces an additional layer of complexity for researchers, who would benefit from greater alignment between platforms in the terminology used for classification granularities—or at minimum, enhanced transparency—to facilitate informed platform selection.

\subsection{Reproducibility}
These classification schemes can bring challenges when trying to reproduce already published results.
Authors’ using them need to be careful to include sufficient information for readers to know which classification scheme has been used, the granularity used when several are available, and which categories have been selected or excluded.
For example saying only that WOS citation topics have been used is not enough, since, as we saw, the difference in granularity between macro-, meso-, and micro-topics is significant.

The other potential obstacle is of course the question of who can access the various platforms.
Researchers are highly dependent on the resources provided by their institution, and not all institutions can, or want to, subscribe to all bibliometric platforms.

\subsection{What about authors’ keywords?}
A concept that we did not mention yet in this analysis is the notion of `author’s keywords'.
These keywords are chosen by the article’ authors to resume the topics covered in their publication.
They are mandatory for all publications in a large number of journals, some journals also require the use of a disciplinary taxonomy from which to select the keywords.\footnote{For example the \href{https://dl.acm.org/ccs}{ACM Computing Classification System} for computer science or the \href{https://www.aeaweb.org/jel/guide/jel.php}{\emph{Journal of Economic Literature} (JEL) classification} for economy.}
They should therefore be useful to construct a thematic collection of publications.
Unfortunately they have a number of limitations.

\begin{enumerate}
    \item In the majority of fields, the lack of normalization in the keywords choices means that the coverture of keywords is hazardous, and a lot of publications can be missed if a collection relies only on authors keywords.
    \item From a technical point of view, this lack of coverage results from the fact that this field is not standardized in the article metadata.
        This means that author’s keywords are not stored in a dedicated, clearly identified metadata field, unlike authors’ names or affiliations.
        This lack of standardization in the technical backbone of the publishing industry has a very concrete impact: authors’ keywords are often not available in bibliometric platforms for filtering publications. % since they are not stored in a dedicated metadata field and are difficult to extract from the publications PDFs.
        Platforms having a limited scope and manual cleaning steps in their data ingestion workflow, such as Web of Science or Scopus, allow filtering on authors keywords to create personalized collections.
        However Platforms relying on automated processes on a larger scope do not index them properly.
\end{enumerate}

\paragraph{Beware of `keywords' fields}
Bibliometric platforms offer several fields named `Keywords'.
We want to emphasize for readers that these fields usually do not correspond to the authors’ keywords listed on a publication.
For example, in The Lens the `Keyword' field lists the keywords indicated on PubMed.

\section{Conclusion and perspectives}
In conclusion, we provided a detailed overview of the classification schemes offered by Web of Science, Scopus, Dimensions, and The Lens, to enable researchers to make informed decisions when selecting a platform for their research.
Our analysis highlights the importance of carefully considering the classification scheme used when constructing a bibliometric collection.
The classification schemes offered by Web of Science, Scopus, Dimensions, and The Lens differ greatly, due to four main factors and strategies implemented by the platforms:
\begin{itemize}
    \item The initial coverage of the platform providing the classification: Small scope manually curated for an emphasis on quality, or large scope automatically harvested for an emphasis on coverage;
    \item The strategy used to assign publications to a category: Experts or artificial intelligence, assignation at journal or document-level, based only on citations or also on the textual content;
    \item The level of granularity of the chosen classification: From very high-level categories to narrow research topics.
    \item The assignment of articles to multiple categories: Whether a publication can be classified into several categories simultaneously, reflecting its disciplinary breadth or its position at the intersection of fields.
\end{itemize}
All these elements have significant implications for research outcomes and the analyses to be performed.
Researchers must be aware of their strengths and weaknesses in order to make the best trade-offs for their needs.
If several classifications appear to meet their needs, researchers should compare them on a small scale (e.g., a pilot study or a subset of their research field) before committing to only one classification and platform for their complete bibliography.
This approach ensures that their research results are accurate and relevant to their research questions.

Looking ahead, several promising avenues emerge for future research.
First, we intend to expand this analysis by incorporating OpenAlex, a rapidly growing open bibliometric platform whose classification system could provide valuable comparative insights.
Second, we plan to conduct a quantitative comparative analysis of these classification schemes to systematically evaluate their coverage, consistency, and potential biases across disciplines.
These perspectives would not only deepen our understanding of bibliometric classification systems but also provide practical guidance for researchers navigating an increasingly complex scholarly information landscape, particularly in interdisciplinary fields where traditional disciplinary boundaries may obscure important contributions.

\section*{Acknowledgment}
The author used Llama 3.3 70B on the French sovereign platform RAGaRenn during the final stage of editing to assist with language refinement and improve the readability of the article.
All text produced with the tool was revised by the author, who takes full responsibility for the final article.

\printbibliography

\end{document}